\RequirePackage[2020-02-02]{latexrelease}
\documentclass[showpacs,showkeys,11pt,
preprint,preprintnumbers,nofootinbib,
groupedaddress,superscriptaddress,amsmath,amssymb]{revtex4}
\usepackage{amsfonts}
\usepackage{amsmath}
\usepackage{graphicx,subfigure}
\usepackage[style=base]{caption}
\usepackage[export]{adjustbox}
\usepackage{epsfig}
\usepackage{url}
\usepackage{multirow}
\usepackage{feynmp}

\newcommand {\be}{\begin{equation}}
\newcommand {\ee}{\end{equation}}
\newcommand {\ba}{\begin{eqnarray}}
\newcommand {\ea}{\end{eqnarray}}

\begin{document}

\title{Observability of triple top quark signal at Future Hadron Colliders}

\author{Ijaz Ahmed}
\email{Ijaz.ahmed@fuuast.edu.pk}
\affiliation{Federal Urdu University of Arts, Science and Technology, Islamabad Pakistan}
\author{Nazima Bi}
\affiliation{Riphah International University, Sector I-14, Hajj Complex, Islamabad Pakistan}
\author{ M. S. Amjad}
\email{sohailamjad@nutech.edu.pk}

\affiliation{National University of Technology (NUTECH), Sector I-12,  Islamabad Pakistan}

\begin{abstract}

The Standard Model cross-sections for the production of single top, top quark in pairs, triple-top quarks, and four top-quarks at three different centre of mass-energies i.e, $\sqrt{s}=$7, 10 and 14 TeV at existing particle colliders as well as at Future Hadron Colliders are studied. A fully kinemtaic analysis with the optimized pre-selection cuts along with invariant mass reconstruction of top quarks is performed for triple-top production processes pp$\rightarrow $tttW,  tttd, tttb in presence of the SM background. All three signal processes are forced to decay in fully hadronic mode. The studies are performed for High Luminosity LHC (HL$-$LHC) at $\sqrt{s}=$14 TeV and for High-Energy LHC (HE$-$LHC) at $\sqrt{s}=$27 TeV. The signal to background ratio and signal significance of all signal and background processes are estimated for both the scenarios. It is found that the the chances of signal observability at HE$-$LHC are higher than that at HL$-$LHC.    
\end{abstract}
\maketitle

\section{Introduction}
Top-quark characteristics are one of the most important aspects of Standard Model (SM). It is also possible that top-quarks will play a key role in breaking electroweak symmetry, which is responsible for the masses of all fundamental particles. Top quarks have some unique properties including enormously large mass, and they could be a gateway to discovery new physics. Also owing to extremely short life time, bare quark properties could be studied as they decay before hadronization. Top-quarks have been observed to be produced singly through weak-interactions and in top-pairs through strong interactions \cite{wicke2010single, gouz1999double}. Due to the large mass, high collision energy is necessary to produce top-quarks. At LHC, a large number of top-quarks were produced at energy of 7 TeV and 8 TeV. As a result, top characteristics have been investigated in great detail and precision. These findings are consistent with the Standard Model's predictions for top quarks (SM). However, in BSM regime, there has been a limited focus on triple top, particularly. For example, \cite {arxiv1710} discusses the triple top in view of searches for scalar bosons. The authors propose that triple-top may be studied in the signature of three leptons plus three $b$-jets, as confirmation. They also argue that triple-top search at HL-LHC could conditionally cover full mass range up to 700 GeV. Also \cite{arxiv1910} includes the discussion on triple top in the context of FCNC induced by the Z` boson. The Z` boson produced in associated with single top quark and decaying to a $t \bar t$ could decay to a triple top final state. This might become dominant as compared to single top under certain coupling conditions. However, there's also the possibility that it remains negligible under different coupling conditions.

Since its discovery \cite{gomez1995observation} in Fermi-lab at Tevatron collider by  the  CDF  and  D0  Collaborations \citep{chakraborty2003top}\citep{barnreuther2012top, abachi1995observation}, the top quark has remained the heaviest elementary particle. It completed the third-generation structure of the Standard Model (SM) and opened the top quark physics area \citep{beneke2000top}.The large mass of the top quark makes phenomenology so important as it is usually the most closely related to new physics proposals Beyond the Standard Model (BSM) \citep{giammanco2016single} . In addition, the it has a very short lifetime and decays without hadronisation \citep{bruscino2020top}, making top quark physics a unique playground to study a bare quark \citep{deliot2020top} . At the Tevatron and Large Hadron Collider (LHC), many properties of the top quark were studied, including the 
 production mechanism, the basic properties such as mass and width, decay \citep{young2020top}. \newline
For both the Fermilab's Tevatron and CERN's Large Hadron Collider (LHC), top-pair and single top production have been extensively studied. Single top production is also of interest, especially for the Standard Model (SM) $V_{tb}$ mixing element, and was observed at the Tevatron \citep{barger2010triple, incandela2009status}. Several search programmes involve top-quark that is commonly considered sensitive to new physics on the TeV scale , e.g. top-quark pair production of opposite-sign or same-sign, single top-quark production, and four top-quark production. Unfortunately, triple top-quark production still does not receive too much attention. Of all the current top-quark-related physics search programs, the triple-top production is very special \cite{cao2019can}. We measure single top, top pair, triple top and four top quark cross-sections at different mass energy centres in this paper and address the discovery potential of triple-top events at the Large Hadron Collider (LHC) and how new physics can significantly affect this channel. 
\section{Single Top Quark Production }
\begin{figure}[ht]
	
	\includegraphics[width=0.20\linewidth]{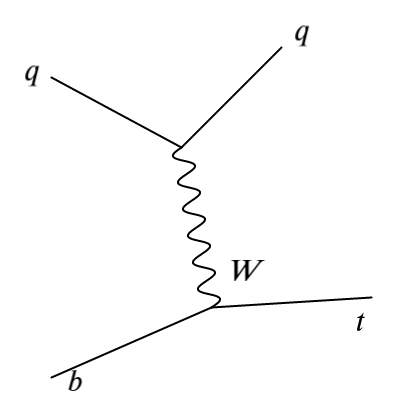}	\hfill
	\includegraphics[width=0.20\linewidth]{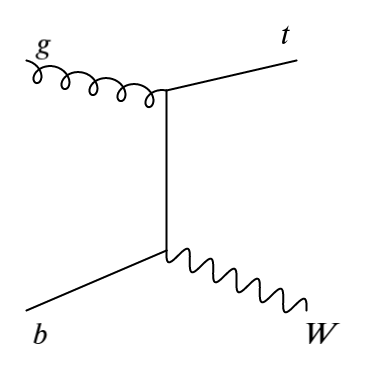}\hfill
	\includegraphics[width=0.25\linewidth]{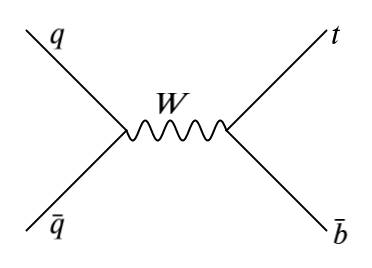}
	\caption{Feynman diagram of single top quark production in SM:  t-channel (left), W-associated production or tW-channel (center) and s-channel (right).}
	\label{fig:feynman4}
\end{figure} 
In Beyond the Standard Model (BSM), single top-quark production is also susceptible to physics including charged Wtb vertex structure, new gauge bosons, new heavy quarks, and top-quark neutral currents that change flavour \citep{tait2000single,atlas2020observation}. There are three different single top quark production modes at Large Hadron Collider (LHC) which are t-channel, s-channel and tW-channel production \citep{bernreuther2008top,kroninger2015top}. Lowest order Feynman diagram for single top quark  production through weak interaction are shown in Fig.\ref{fig:feynman4} .
\begin{table}[ht]
	\centering
	
		\begin{tabular}{|c|c|c|c|c|}
			\hline
			Process & No. of   &  $ \surd s $ = 7 TeV  &  $ \surd s $ = 10 TeV  &  $ \surd s $ = 14 TeV  \\ 
			
			 & Diagrams & & & \\ \hline
			$\sigma(t-channel)$[fb]	 &4  &$ 1.8039\times10^{4} $   &$3.5832\times10^{4} $ & $ 4.9552\times10^{4} $ \\
			\hline
			$\sigma	(s-channel)$[fb] &4 & $ 9.2312\times10^{2} $  & $ 1.5027\times10^{3} $ &  $ 2.3140\times10^{3} $  \\
			\hline
			$\sigma(tW-channel)	$[fb] &4 & $ 2.9856\times10^{3} $ & $
			6.5386\times10^{3} $  & $ 1.5074\times10^{4} $	 \\ \hline
			
		\end{tabular}
		\label{tab:Table.No.1} 
	\caption{Cross section for single top quark production through weak interaction at Large Hadron Collider (LHC).}
\end{table} 
  \begin{figure}[ht]
   	\centering
   	{\Large \textbf{\includegraphics[width=0.5\linewidth]{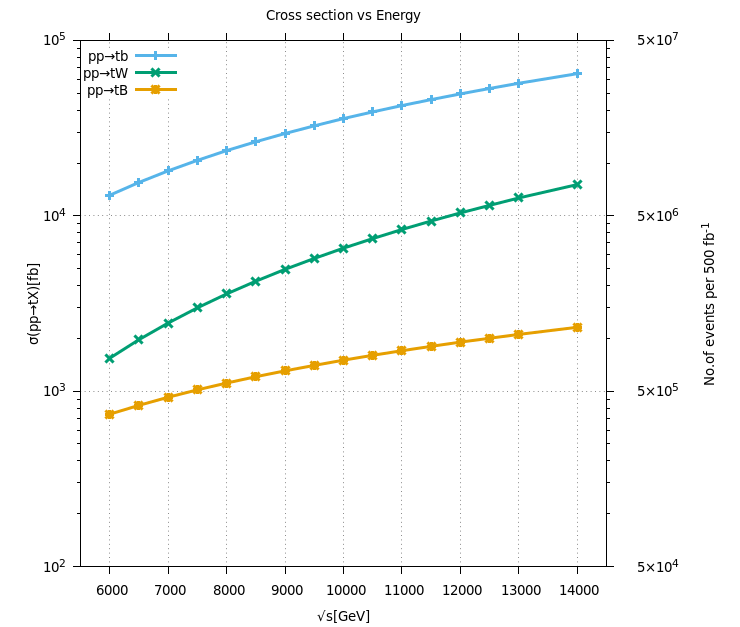}}}
   	\caption{Cross section for single top production in the Standard Model for different LHC center of mass energies.}
   	\label{fig:st}
   \end{figure}
\section{Top Pair Production}
\begin{figure}[ht]
	\includegraphics[width=0.26\linewidth]{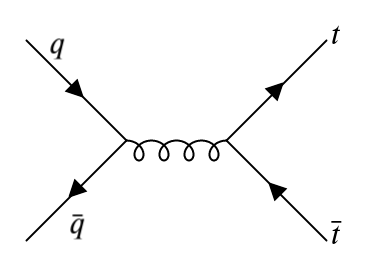}\hfill
	\includegraphics[width=0.26\linewidth]{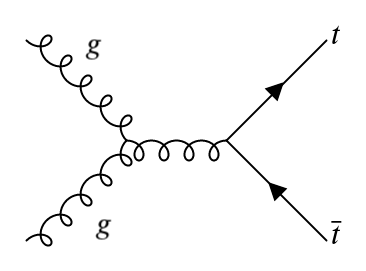}\hfill
	\includegraphics[width=0.26\linewidth]{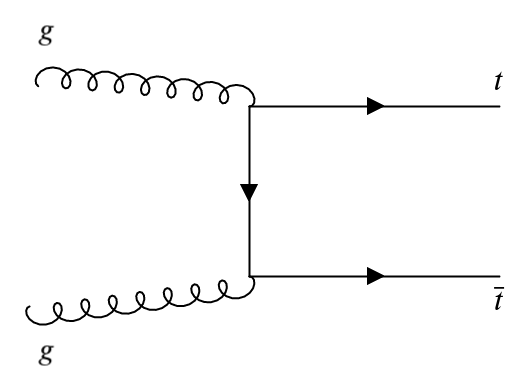}\hfill
	\includegraphics[width=0.22\linewidth]{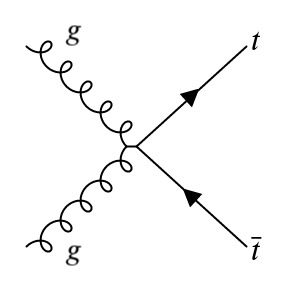}
	\caption{Lowest order Feynman diagrams contributing to top quark pair production at Large Hadron Collider (LHC).}
	\label{fig:img24}
\end{figure}
The production of triple top quarks is a mixture of single top quarks and pairs \citep{barger2010triple}. According to the SM, the strong interactions generate top pairs at the Tevatron, as well as at the LHC \citep{incandela2009status} and single top quark production mostly through electroweak interaction with W-boson \citep{bernreuther2008top, abazov2009observation}.  Top -quark pairs are formed by quark-antiquark $(q\bar{q}\rightarrow t\bar{t})$ annihilation and gluon-gluon fusion $(gg\rightarrow t\bar{t})$. In the Tevatron collider, the first is the most dominant, while in the LHC, the second is dominant \citep{beneke2000top,han2008top,bernreuther2008top}. Fig.\ref{fig:img24} shows the Faynman diagram of the lowest order for top pair production by strong interaction. \
 \begin{table}[ht]
	\centering
	
		\large
		\begin{tabular}{ |c|c |c |c |c|}
			\hline
			
			Process &No. of  &  $ \surd s $ =7 $ TeV $ &  $ \surd s $ =10 $ TeV $ &  $ \surd s $ =14 $ TeV $ \\ 
			& Feynman Diagrams & & & \\ \hline
	
			$\sigma(q\bar{q}\rightarrow t\bar{t})$[fb] &38 & $ 5.6691\times10^{3} $  & $ 1.0336\times10^{4} $ & $ 1.7143\times10^{4} $ \\
			\hline
			$ \sigma(gg\rightarrow t\bar{t})$[fb] &3 & $ 4.8952\times10^{4} $  & $ 1.4287\times10^{5} $ &  $ 3.5331\times10^{5} $  \\
			
			\hline
		\end{tabular} 
		\caption{Cross section for top quark pair production through strong interaction at LHC }
		\label{tab:Table.2}
\end{table}\newline
We can see from Table \ref{tab:Table.2} that the production of top quark pairs in LHC is much greater than that of Tevatron because of higher collision energy and higher luminnosity \citep{han2008top}.
\begin{figure}[ht]
	\centering
	\includegraphics[width=0.6\linewidth]{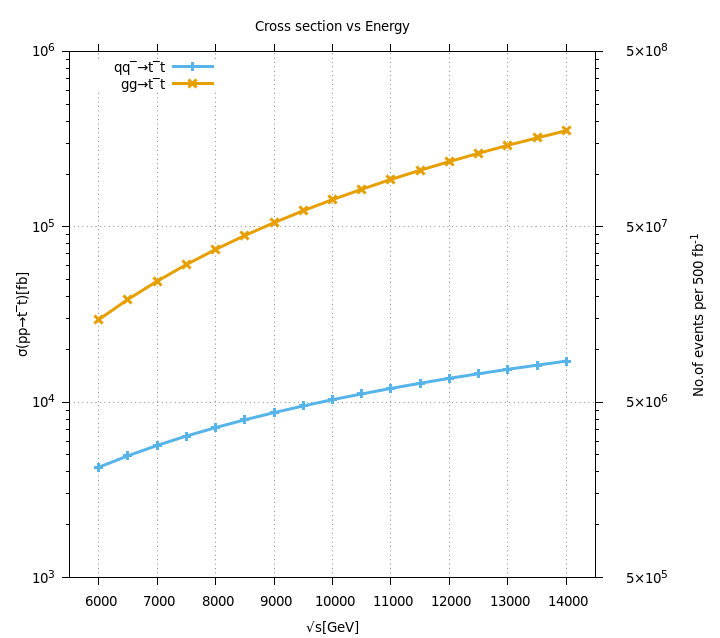}
	\caption{Top quark pair production cross section for quark anti-quark annihilation and gluon-gluon fusion processes at Large Hadron Collider (LHC).}
	\label{fig:tp}
	\end{figure}\newpage
\section{Triple Top Quark Production}
\begin{figure}[ht]
 \centering
 
        \includegraphics[width=0.3\linewidth]{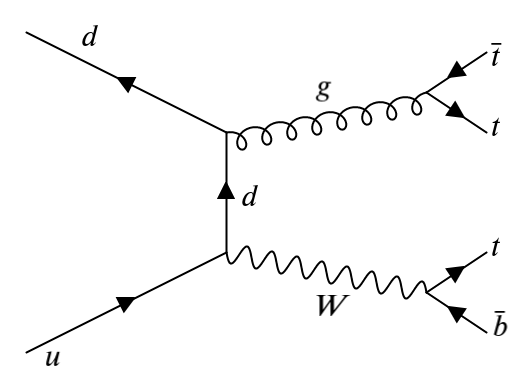}	\hfill
	\includegraphics[width=0.3\linewidth]{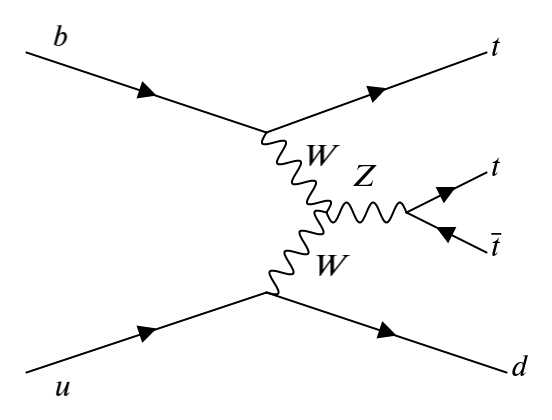}\hfill
	\includegraphics[width=0.31\linewidth]{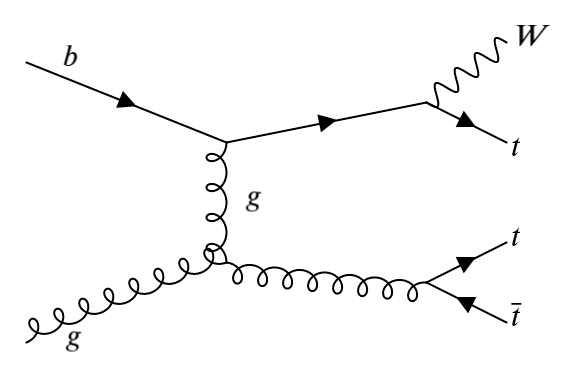}
 
	\caption{Feynman diagrams for triple top quark production in Standard Model (SM) at Large Hadron Collider (LHC) corresponding to three processes.}
	
	\label{fig:feynman13}
\end{figure}
 Triple top quark has three different production modes in the Standard Model (SM) at LHC \citep{boos2021triple}. The three production modes of triple top quark are $pp\rightarrow 3t+W^{\pm}$, $pp\rightarrow 3t+\bar{b}$ and $pp\rightarrow 3t+d$. \citep{barger2010triple}.
\begin{table}[ht]
	\centering
		\begin{tabular}{ |c|c|c|c |c|}
			\hline
			Process & No. of&  $ \surd s $ = 7 $ TeV $ &  $ \surd s $ = 10 $TeV $ &  $ \surd s $ = 14 $ TeV $ \\
			& Diagrams & & & \\ \hline
			$\sigma(pp\rightarrow tt\bar{t}+W^{-})$[fb] & 118 & $ 1.8039\times10^{4} $  & $ 3.5832\times10^{4} $ & $ 4.9552\times10^{4} $ \\
			\hline
			$\sigma(pp\rightarrow tt\bar{t}+d)$[fb] &76& $ 9.2312\times10^{2} $  & $ 1.5027\times10^{3} $ &  $ 2.3140\times10^{3} $  \\
			\hline
			$\sigma(pp\rightarrow tt\bar{t}+\bar{b})$[fb]&36 & $ 2.9856\times10^{3} $ & $
			6.5386\times10^{3} $  & $ 1.5074\times10^{4} $	 \\
			\hline
			
		\end{tabular}
	 	\caption{Cross section for triple top quark production at LHC.}
	 	\label{tab:Table.3}
\end{table} 
Table \ref{tab:Table.3} shows that triple top quark cross section is low. Due to its small cross section, triple top quark production is very rare \citep{khanpour2020probing}. 
\begin{figure}[!bh]
	\centering
	\caption{Cross section for triple top production in SM for different LHC center of mass energies.}
	\label{fig:ttp}
\end{figure}
 \newpage 
 The triple top cross section is very small compared to the Standard Model (SM) prediction which makes it an interesting channel to study. The process $pp\rightarrow tt\bar{t}+W^{-}$ has the larger cross-section as compared to other processes at different center of mass energies. 
 \section{Four Top Quark Production}
 \begin{figure}[ht]

	\includegraphics[width=0.4\linewidth]{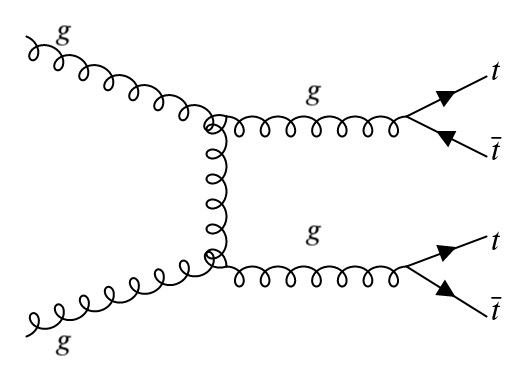}\hfill
	\includegraphics[width=0.4\linewidth]{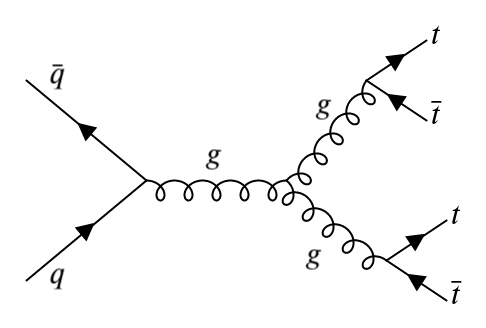}
	
 	\caption{Feynman diagrams for four top quark production in Standard Model (SM) at Large Hadron Collider (LHC) corresponding to quark anti-quark annihilation and gluon fusion  processes.}
 	\label{fig:img27}
 \end{figure}
 Four top quarks has two production modes quark anti-quark annihilation and gluon-gluon fusion. The gluon-gluon fusion process is dominant at LHC. The gluon-gluon fusion has a contribution of $90 \%$ and quark anti-quark annihilation of $10 \%$. 
 \begin{table}[ht]
 	\centering
 
 		\large
 		\begin{tabular}{ |c| c |c |c |c | }
 			\hline
 			
 			Process & No. of  &  $ \surd s $ = 7 $ TeV $ &  $ \surd s $ = 10 $ TeV $ &  $ \surd s $ = 14 $ TeV $ \\ 
 			& Diagrams & & & \\ \hline
 			
 			$\sigma(q\bar{q}\rightarrow t\bar{t}t\bar{t})$[fb] & 54 & $ 5.6691\times10^{3} $  & $ 1.0336E\times10^{4} $ & $ 1.7143\times10^{4} $ \\
 			\hline
 			$\sigma(gg\rightarrow t\bar{t}t\bar{t})$[fb] &54  & $ 4.8952\times10^{4} $  & $ 1.4287\times10^{5} $ &  $ 3.5331\times10^{5} $  \\		
 			\hline
 		\end{tabular} 
 	 \caption{Cross section for four top quark  production at Large Hadron Collider (LHC). }
 	 \label{tab:Table.4}
 \end{table}
The cross-section of four top quarks production is around five orders of magnitude smaller than the top pair production and so it has yet to be observed \citep{cms2017search}. 

 Table \ref{tab:Table.4} shows the cross section for quark anti-quark annihilation $(q\bar{q}\rightarrow t\bar{t}t\bar{t})$ and
 gluon-gluon fusion $(gg\rightarrow t\bar{t}t\bar{t})$ at different center of mass energy for four top quark  production.
 
 \begin{figure}[!ht]
 	\centering
 	\includegraphics[width=8cm, height=6cm]{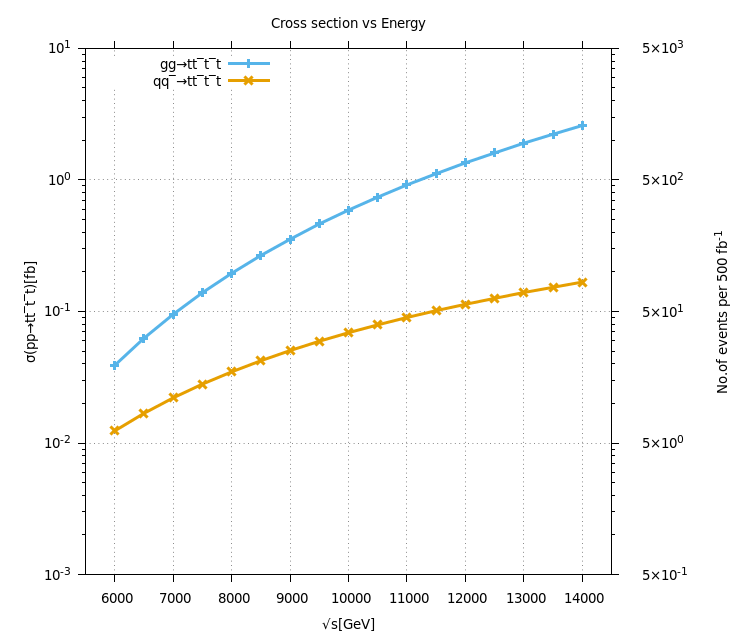}
 	\caption{Cross section for four top production in the Standard Model (SM) for different Large Hadron Collider (LHC) center of mass energies.}
 \end{figure}

\begin{figure}[!ht]
	\centering
	\includegraphics[width=8cm, height=6cm]{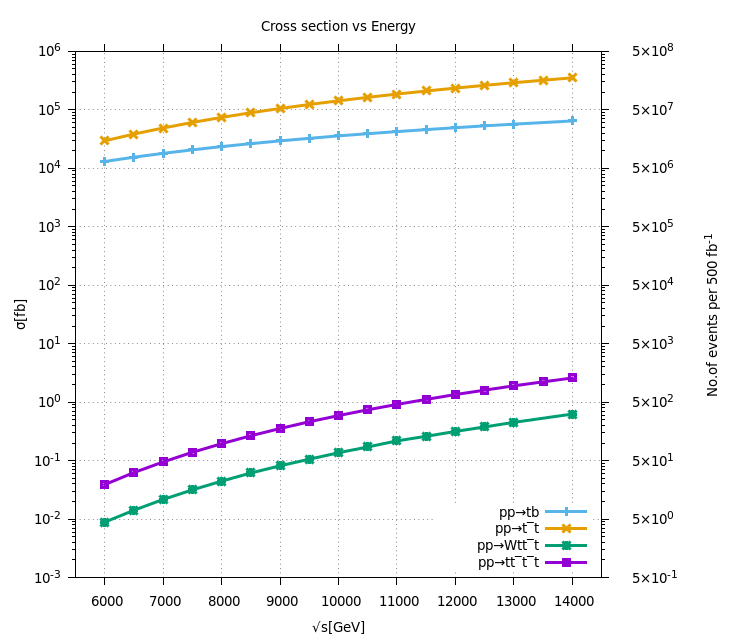}
	\caption{Cross section for single, pair, triple and four top production in the (SM) for different LHC center of mass energies.}
\end{figure}

 \section{Signal and Background processes} 
In this study, various scattering mechanisms are used as signals. The $tt\bar{t}W, tt\bar{t}\bar{b}$, and $tt\bar{t}d$ are scattering processes. All of these scattering mechanisms are generated by proton-proton collisions at $\sqrt{s}$=14 TeV. In all of the triple-top signal processes, hadronic decay of the W$-$boson is taken into account. In the $tt\bar{t}W, tt\bar{t}\bar{b},$ and $tt\bar{t}d$ scattering processes, top-quark (t) decays into W$-$boson and bottom quark (b). W decays to produce light jets (u, d, c, s). As a result, there are eleven jets in the final state of the scattering process $tt\bar{t}W$, consisting of eight light jets and three b$-$jets. $pp\rightarrow tt\bar{t}W \rightarrow 4W^{\pm}+3bjets \rightarrow 8jets+3bjets$. There are ten jets in the final state of the scattering process $tt\bar{t}B$, which includes six light jets and four b-quarks jets $pp\rightarrow tt\bar{t}\bar{b} \rightarrow 3W^{\pm}+4bjets \rightarrow 6jets+4bjet$. Similarly, the $tt\bar{t}d$ scattering process produces ten jets, seven of which are light jets and three of which are b-jets $pp\rightarrow tt\bar{t}d \rightarrow 3W^{\pm}+3bjets \rightarrow 7jets+3bjets$.\\
The following background processes of SM with similar final state topologies are produced while analyzing these signal processes. In $t\bar{t}Z, t\bar{t}W^{\pm} and W^{+}W^{-}Z$, top quark decays into W boson and b jet. When Z and W decay, they produce a pair of light jets. In $pp\rightarrow t\bar{t}Z\rightarrow$ 6jets+2bjets, $t\bar{t}W\rightarrow$ 6jets+2bjets and in $W^{+}W^{-}Z\rightarrow 6jets$.
\begin{table}[!ht]
	\centering
	
		\begin{tabular}{|c|c|c|c|c| c| c| c|c|c| }
		\hline
		\bfseries	$\sqrt{s}$  [TeV]& $tt\bar{t}W$ &$t\bar{t}tb$ & $t\bar{t}td$ &  $t\bar{t}h$& $t\bar{t}Z$ & $t\bar{t}W$ & tWZ & WWZ & tZj   \\[2ex] 
        [TeV]& fb &fb &fb &fb &fb &fb &fb & fb&fb \\
		\hline
		
		14    & 1.5	 & 0.1   & 0.2  & 483    &703      & 402      & 141 & 90 & 816   \\[1.8ex]
		\hline
		27    &12.7     &0.45    &1.2    & 838 & 3329 & 1176 & 691 & 263 & 3142 \\[1.8ex]
		\hline
		100 & 352  &3.5    &13    & 11490 & 44800 & 6976 & 8970 & 1463 & 26380 \\[1.8ex]
		\hline
					
	\end{tabular}
	 
	\caption{Cross-sections of three signals and background 
 processes are shown at three different energies using Next-to-Next Leading Order parton density function }
\end{table} 
\section{Event Selection and Collider Analysis}
 A complete analysis of three processes whose cross sections are presented in above section are studied. Three signal processes are explored where three top quarks are produced along with additional particle in each process and fully hadronic decay modes are selected. The parton density function is provided by LHAPDF 5.9.1 \cite{whalley2005houches} with the version CTEQ 6.6. The background processes $W^{+}W^{-}Z$, $t\bar{t}W^{\pm}$, $t\bar{t}Z$ are generated with calchep \cite{belyaev2013calchep} with a kinematic preselection cut applied on jets as $E_{T}^{jets}>15$ GeV and $|\eta|<3.0$, \cite{sjostrand2008brief}. All the signal processes are also produced with Calchep. The output of both these packages is in Les Houches Event  (LHE) format is used by PYTHIA8 for parton showering, gluon radiation, fragmentation and hadronization. PYTHIA calculates their relative efficiency as well. The HepMC v2.06.09 interface with PYTHIA is then used to record events. FastJet v3.3.4 is then coupled with PYTHIA for jet definition and reconstruction. The jet cone size is fixed at $\Delta R=0.4$, where $\Delta R=\sqrt {(\Delta \eta)^{2}+(\Delta \phi)^{2}}$ is jet cone radius, $\phi$ is azimuthal angle and $\eta =-ln$tan$\theta/2$ is pseudorapidity. The output is then analysed with ROOT v6.14.06.
 In this study, various scattering mechanisms are used as signals. The $tt\bar{t}W, tt\bar{t}\bar{b}$, and $tt\bar{t}d$ are scattering processes.
All of these scattering mechanisms are generated by p-p collisions at $\sqrt{s}$=14 TeV High Luminosity LHC (HL-LHC) and $\sqrt{s}$=27 TeV center-of-mass energy at High Energy LHC (HE-LHC). In all of the triple-top signal processes, a hadronic decay of the W$-$boson is taken into account. 

Various selection cuts are used to reduce background while keeping the signal. The chi-square method is used to reconstruct the physics objects in which we are interested. Several kinematic variables are plotted during entire analysis to examine the distributions of W bosons and top quark etc. 

Jets are reconstructed using the Anti$-k_{t}$ technique with R and $ \Delta R$ set to 0.4 in this investigation.\\
Some of the jets in this mechanism extend beyond the cone size due to a variety of causes such as detector impairment, magnetic field influence, and material influence. All of these jets are sorted by $p_T$ and the following kinematic cut on jets is applied.

\begin{center}
$P_{T}^{ljets} \geq$ 15 GeV and $\mid\eta\mid\leq$ 2.5
\end{center}  
\medskip

The selected jets are then tagged as either a b-jet or a light jet. In order to to this, we do $\Delta R$ matching of jets with the parton level b- and c-quarks. Charm quarks are used to increase efficiency since their masses are closer to those of bottom quarks.  The jets which are within $\Delta R < 0.2$ are identified as b-jets and all the other jets which are farther away from b-quarks are identified as light jets. Once the tagging is done, we apply a multiplicity cut on jets. In $tt\bar{t}W$ analysis, there are eight light jets and three b jets so we only keep the events where we have

\begin{center}
$N_{ljets}\ge8$, $N_{bjets}\ge3$
\end{center}

\begin{figure}[!ht]
	\centering
	\includegraphics[width=10cm, height=7cm]{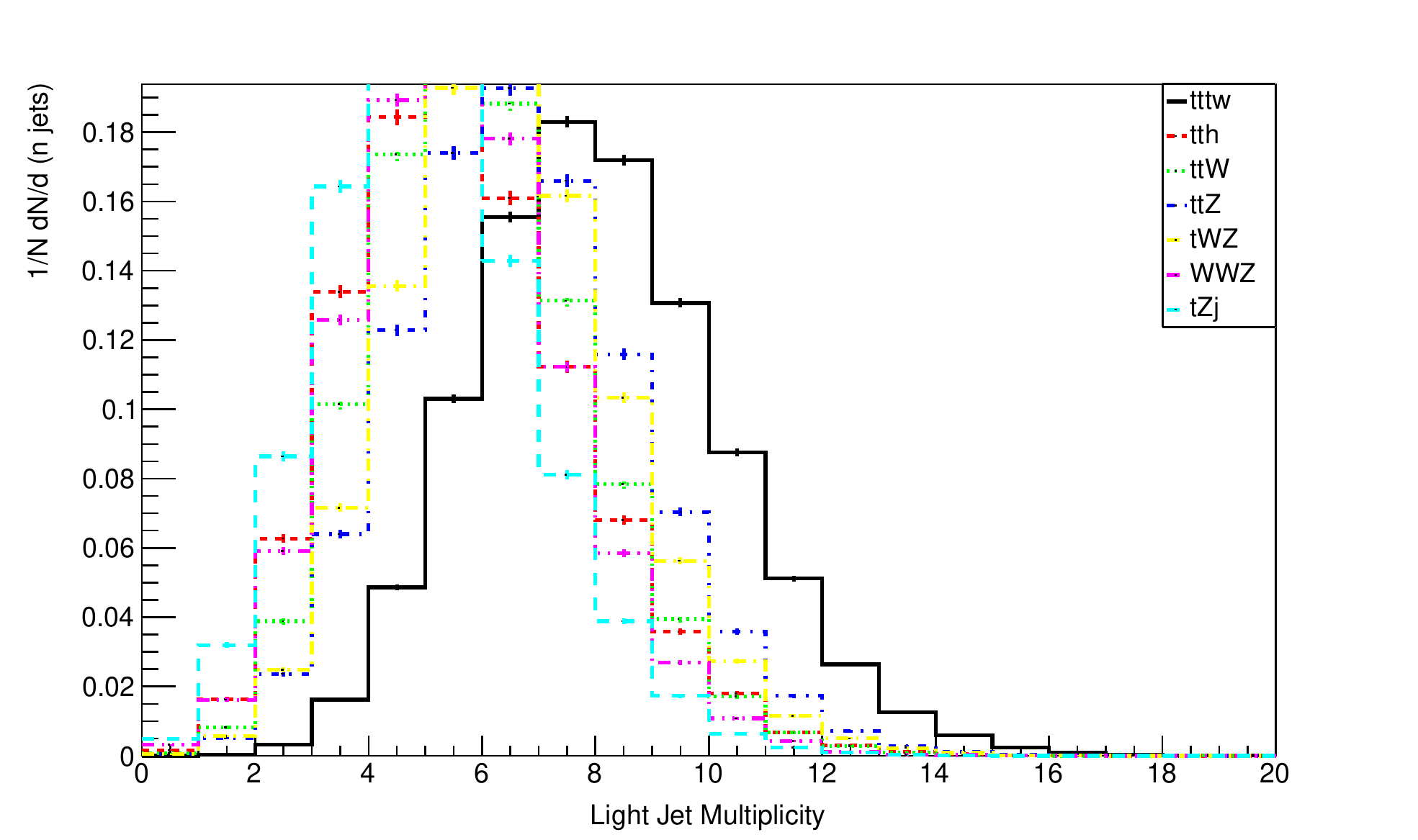}
	\caption{The jet multiplicity distributions of both signal and background events are shown at $\sqrt{s}$=14 TeV}
	\label{fig:4tc}
\end{figure}
\begin{figure}[!ht]
	\centering
	\includegraphics[width=10cm, height=7cm]{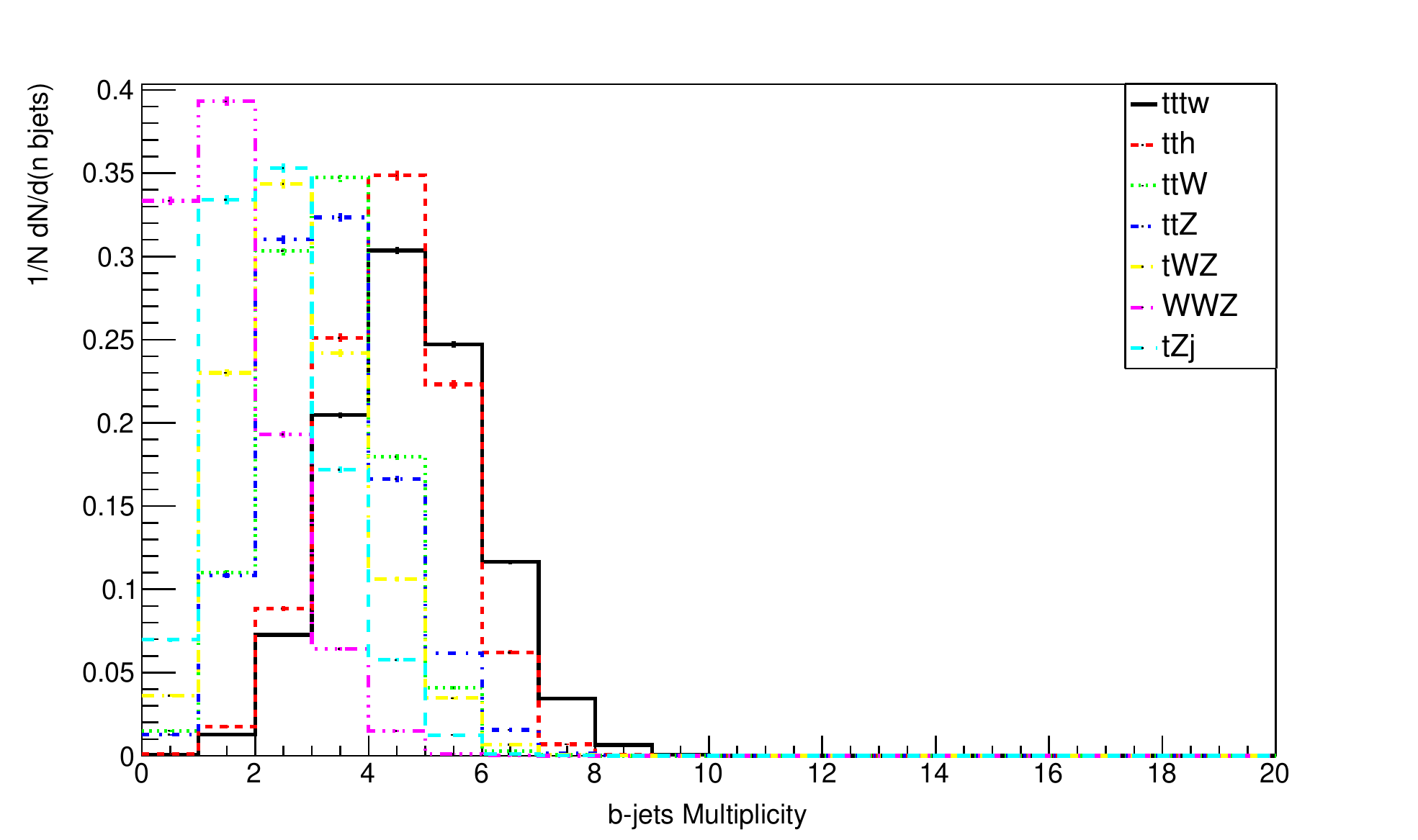}
	\caption{The b-jets multiplicity distributions in both signal and background events}
	\label{fig:4tcused}
\end{figure}

\begin{figure}[!ht]
	\centering
	\includegraphics[width=10cm, height=7cm]{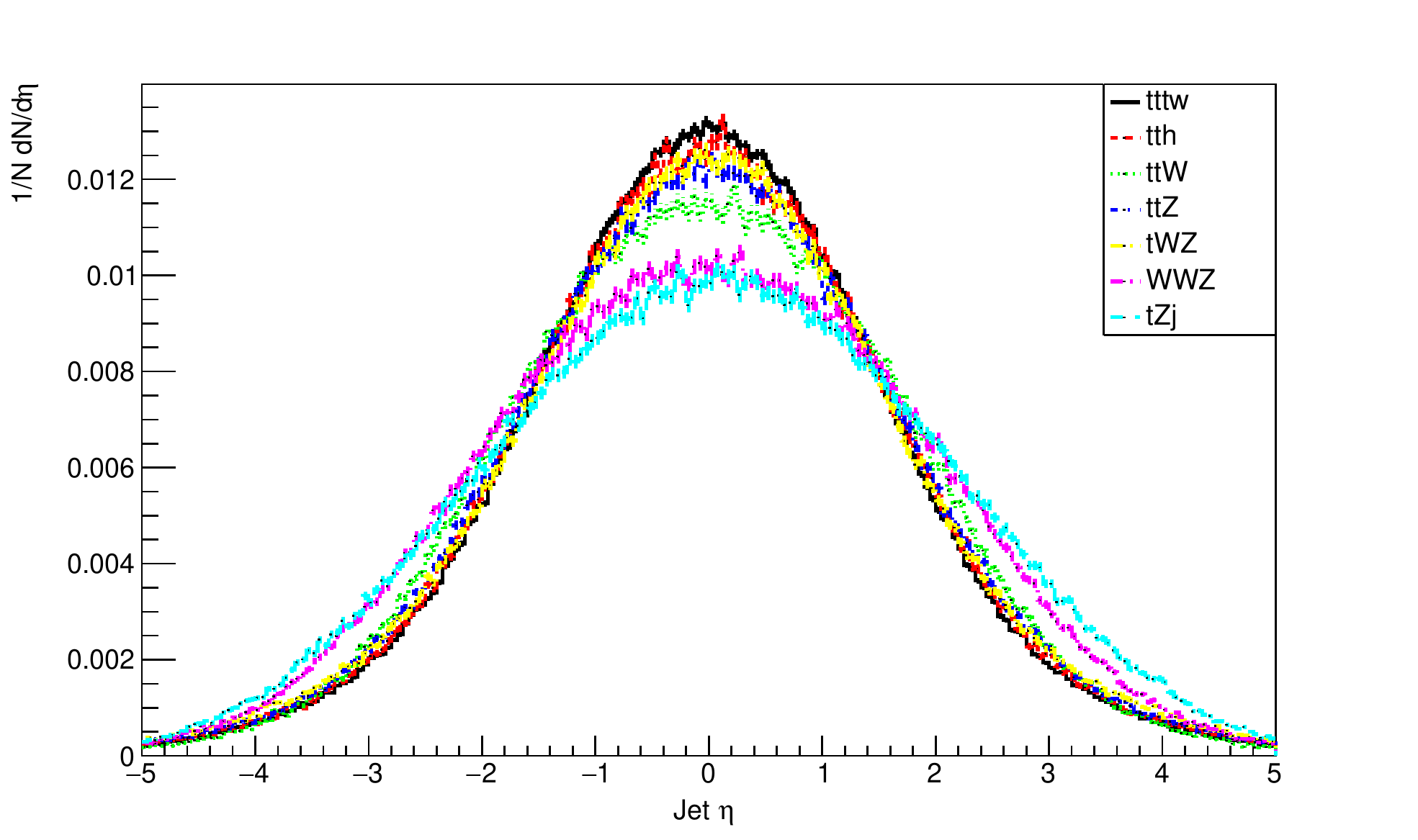}
	\caption{Pseudorapidity distributions of signal and background selected jets at $\sqrt{s}$=14 TeV}
	\label{fig:4tceta}
\end{figure}

\begin{figure}[!ht]
	\centering
	\includegraphics[width=10cm, height=7cm]{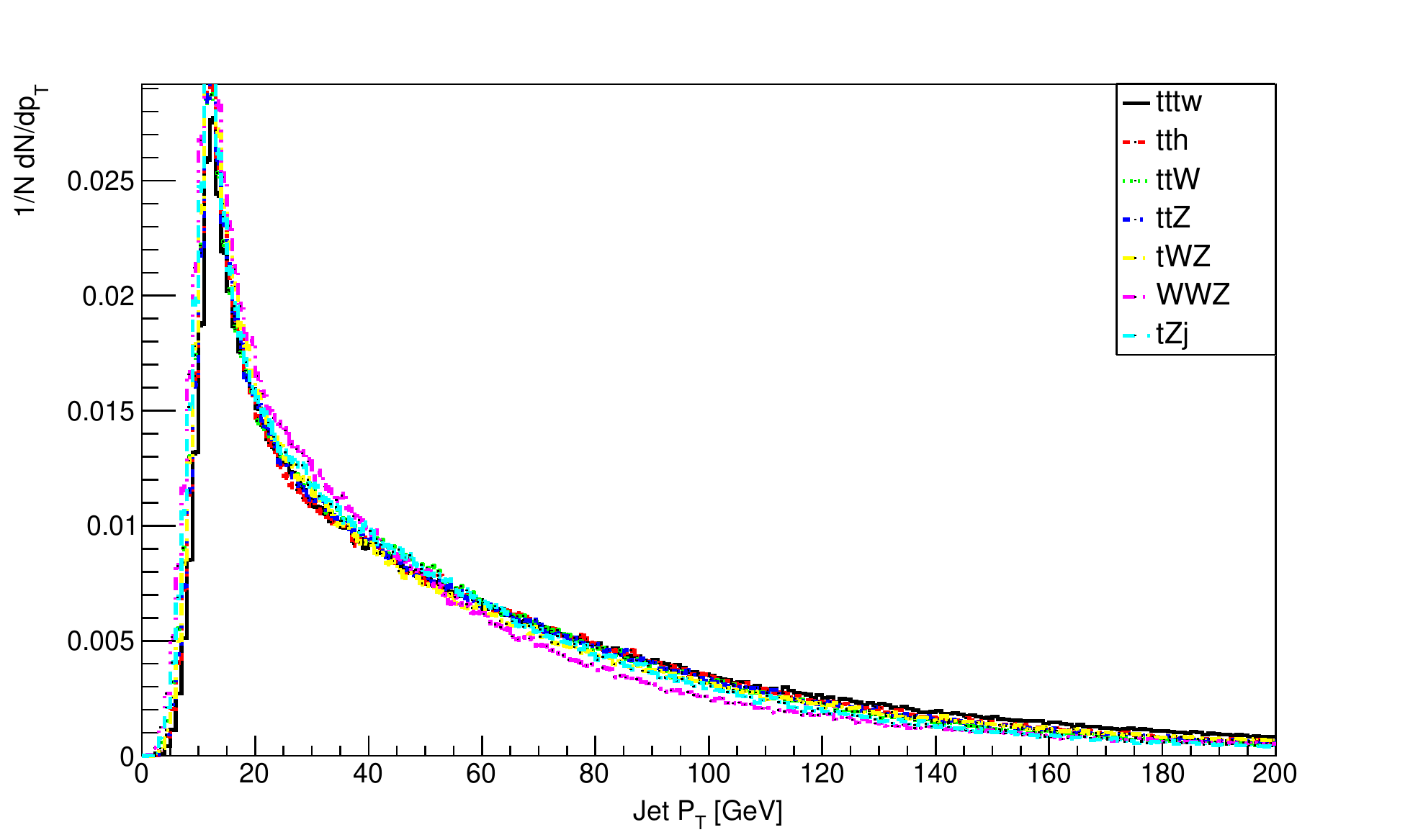}
	\caption{Transverse momentum distributions of selected jets at $\sqrt{s}$=14 TeV}
	\label{fig:4tcpt}
\end{figure}

Similarly, for $tt\bar{t}\bar{b}$, there are six light jets and for b jets so event selection cut becomes
\begin{center}
$N_{ljets}\ge6$, $N_{bjets}\ge4$
\end{center}

and for process $tt\bar{t}d$

\begin{center}
$N_{ljets}\ge7$, $N_{bjets}\ge3$
\end{center}
In figure \ref{fig:4tc} light jet multiplicity of signal processes with backgrounds are given. It can be seen almost 20 percent jets are passed through jets kinematic cut. Also it can be seen jet multiplicity of scattering processes is almost same. Distribution of pseudorapidity and transverse momentum of selected jets are also given in figure \ref{fig:4tceta} and figure \ref{fig:4tcpt} respectively  .

It can be seen from figure, prominent signals with b-jets. In case of $tt\bar{t}\bar{b}$ scattering process four b-jets efficiency is in range of 50 percent. But in case of
$tt\bar{t}d$ and $tt\bar{t}W$ it increases upto 75 percent to 80 percent. Presence of SM backgrounds as compare to signal is low.
Figure \ref{fig:4tcused} shows the distributions of b-jet multiplicities for various signals and backgrounds. 
\section{Reconstruction of invariant masses}
In particle physics, the invariant mass is defined as it's mass in the rest frame and is given as
\begin{equation}
m_{0}c^{2}=(\frac{E}{c})^{2}-\mid p \mid^{2}
\end{equation} 
Set c = 1 for convenience,
\begin{equation}
m_{0}=({E})^{2}-\mid p \mid^{2}
\end{equation} 
After selecting desired events from randomly produced events. The reconstruction of invariant masses from decays product is the next step. First, the invariant mass of the W-boson is reconstructed. For this, all events having at least 8 jets for the $tt\bar{t}W$ signal, 6 jets for the $tt\bar{t}\bar{b}$ signal, and 7 jets for the $tt\bar{t}d$ signal are chosen, and the invariant mass is reconstructed using all possible light jets pairs using formula.
\begin{equation}
\mathbf{ W_{j_{1}j_{2}}=\sqrt{(E_{j_{1}}+E_{j_{2}})^{2}-(p_{x_{j_{1}}}+p_{x_{j_{2}}})^{2}-(p_{y_{j_{1}}}+p_{y_{j_{2}}})^{2}-(p_{z_{j_{1}}}+p_{z_{j_{2}}})^{2}}}
\end{equation}
\section{Chi-square method}
The chi-square method is then used to reconstruct the physics object of interest with correct invariant mass. For reconstruction of W boson, we check the invariant mass of di-jet objects and chose the six jets that give the minimal value of chi-square defined as,

\begin{equation}
\chi_W^{2}=\sum_{i=1}^{3}\left(\dfrac{m_{i,jj}-m_{W}}{\sigma_{m,jj}}\right) ^{2}
\end{equation}

where $m_{i,jj}$ is the di-jet mass, $m_{W}$ is the mass of $W$ boson and ${\sigma_{m,jj}}$ is the width of the di-jet mass distribution obtained from the parton matched jets. The events are only selected if the $\chi_{W,min}^{2} < 10$. Once the W's are reconstructed, we again apply the same method to reconstruct top quark candidates using the selected b-jets and the reconstructed W's.

\begin{equation}
\chi_t^{2}=\sum_{i=1}^{3}\left(\dfrac{m_{i,jjb}-m_{t}}{\sigma_{m,jjb}}\right) ^{2}
\end{equation}

where the light jets are the ones chosen to make W bosons, $m_{i,jjb}$ is a tri-jet mass,  $m_{t}$ is the mass of top quark and ${\sigma_{m,jjb}}$ is the width of tri-jet distribution. Once again, the event is only selected if  $\chi_{t,min}^{2} < 10$.

\section{Event selection efficiencies}
In this study, 80000 events are generated and combined for signal processing in order to improve simulation results on selected kinematical cuts. Then signal efficiency corresponding to each selection cut is computed and then at the end, total efficiency is calculated. 
\begin{table}[!ht]
	\centering
	
		\begin{tabular}{ |c|c|c |c|c| c| c| c|}
		\hline
		\bfseries	Process&$tt\bar{t}W$ &$t\bar{t}Z$ &  $ t\bar{t}W$ & $t\bar{t}H$ & tWZ & tZj & WWZ  \\[2ex] 
		\hline
		\hline
		$N_{ljets}\leq8$       &0.49     &0.25    &0.14    & 0.13 & 0.20 & 0.06 & 0.10  \\[1.8ex]
		\hline
		$N_{bjets}\leq3$       &0.86     &0.38    &0.34    & 0.76 & 0.20 & 0.08 & 0.017\\[1.8ex]
		\hline
		$\chi^{2}_{W, min}<10$ &0.87     &0.87    &0.85    & 0.86 & 0.85 & 0.80 & 0.78 \\[1.8ex]
		\hline
		$\chi^{2}_{t, min}<10$ & 0.89    &0.85    &0.89    & 0.91 & 0.78 & 0.81 & 0.62\\[1.8ex]
		\hline
		Total Efficiency       &0.32     &0.07   &0.037    & 0.08 & 0.02 & 0.003 &0.001 \\[1.8ex]
  \hline
		K factor   &  -& 1.27  & 1.24 & 1.21    &	- & -  &  -       \\[1.8ex]
		\hline
Total Efficiency       &0.32     &0.09   &0.045    & 0.097 & 0.02 & 0.003 &0.001\\[1.8ex]
		\hline
			
	\end{tabular}
	 
	\caption{Efficiencies of signal $tt\bar{t}W$ and  SM background process efficiencies at various kinematics and selection cuts.}
	\label{tab:Table.5}
\end{table} 
Total efficiency corresponds to nine jet final states obtained at the end of the simulation, which consists of six light jets coming from three W-boson decay and b-jets resulting from top-quark decay.  All these efficiency are calculated at centre of mass energy 14 TeV . All these efficiencies are mentioned in Tabs \ref{tab:Table.5}, \ref{tab:Table.6} and \ref{tab:Table.7}. The QCD k-factor values of $t\bar{t}Z$, $t\bar{t}W$, $t\bar{t}H$ are written in table \ref{tab:Table.5}, which are calculated by taking the ratio of $\sigma_{NLO} over \sigma_{LO}$ of given process reported in \cite{KFttZ}, \cite{KFttW} and \cite{KFttH} respectively.
\begin{table}[!ht]
	\centering
		\begin{tabular}{ |c|c|c |c|c| c| c| c| }
		\hline
		\bfseries	Process&$tt\bar{t}\bar{b}$ &$t\bar{t}Z$ &  $ t\bar{t}W$ & $t\bar{t}H$ & tWZ & tZj & WWZ   \\[2ex] 
		\hline
		\hline
		$N_{ljets}\leq6$       &0.61     &0.62    &0.48    & 0.41 & 0.58 & 0.30   & 0.40\\[1.8ex]
		\hline
		$N_{bjets}\leq4$       &0.60     &0.14    &0.12    & 0.48 & 0.066 & 0.014 & 0.003\\[1.8ex]
		\hline
		$\chi^{2}_{W, min}<10$ &0.87     &0.86    &0.84    & 0.86 & 0.83 & 0.78   & 0.74\\[1.8ex]
		\hline
		$\chi^{2}_{t, min}<10$ &0.84     &0.83    &0.87    & 0.88 & 0.79 & 0.75   & 0.75\\[1.8ex]
		\hline
		Total Efficiency       &0.27     &0.063   &0.041   & 0.15 & 0.025 & 0.003 & 0.001\\[1.8ex]
		\hline
	K factor   &  -& 1.27  & 1.24 & 1.21    &	- &  - &  -       \\[1.8ex]
		\hline
Total Efficiency        &0.27     &0.08   &0.050   & 0.18 & 0.025 & 0.003 & 0.001\\[1.8ex]
		\hline
			
	\end{tabular}
 
	\caption{Signal $tt\bar{t}\bar{b}$ SM background process efficiencies at various kinematics and selection cuts.}
		\label{tab:Table.6}
\end{table}  
It can be seen that the efficiency of light jets for various kinematical cuts ranges from 15 percent to 26 percent for signals. In addition, the efficiency of b-jet production ranges from 50 percent to 80 percent for signals. The total efficiency of top-mass reconstruction and nine jet final states for the specified signal situation is very low, ranging from 6 percent to 8 percent. 
\begin{table}[!ht]
	\centering
	
		\begin{tabular}{ |c|c|c |c|c| c| c| c| }
		\hline
		\bfseries	Process&$tt\bar{t}d$ &$t\bar{t}Z$ &  $ t\bar{t}W$ & $t\bar{t}H$ & tWZ & tZj & WWZ   \\[2ex] 
		\hline
		\hline
		$N_{ljets}\leq7$    &0.47	 & 0.42   & 0.28  &  0.26    &  0.38    &  0.15      & 0.22    \\[1.8ex]
		\hline
		$N_{bjets}\leq3$    &0.81	 & 0.40   & 0.37  &  0.78    &  0.22    &  0.088      &0.02    \\[1.8ex]
		\hline
     	$\chi^{2}_{W, min}<10$  &0.96	 & 0.97   & 0.97  &  0.98    &  0.96    &  0.95      & 0.93    \\[1.8ex]
		\hline
		$\chi^{2}_{t, min}<10$       &0.706	 & 0.69   & 0.76  &  0.79    &  0.60    &  0.56      & 0.55    \\[1.8ex]
		\hline
		Total Efficiency    &0.26	 & 0.12   & 0.081 &  0.16    &  0.049   &  0.007      & 0.002    \\[1.8ex]
		\hline
		K factor   &  -& 1.27  & 1.24 & 1.21    &	- & -  & -    \\[1.8ex]
		\hline
Total Efficiency   &0.26	 & 0.15   & 0.1 &  0.19    &  0.049   &  0.007      & 0.002\\[1.8ex]
		\hline

	\end{tabular}
	 
	\caption{Signal $tt\bar{t}d$ and SM background process efficiencies at various kinematics and selection cuts.}
	\label{tab:Table.7}
\end{table}
\section{Signal significance}
To test the observability of the triple top mass at various kinematical cuts, signal significance is calculated for each triple top-mass distribution shown in the figures from 6.31 to 6.39 by incorporating the total number of signal and background candidate masses within the selected mass limit.
Signal significance is calculated using integrated luminosity 3000 $fb^{-1}$. The computed results, which include signal S and background B of candidate masses, signal to background ratio S/B, and signal significance $S/\sqrt{B}$ at $\sqrt{s}$ = 14 TeV and 27 TeV, are shown in the table \ref{significance}, \ref{significance1}.
\begin{figure}[!ht]
	\centering
	\caption{Candidate mass distribution of triple top at integrated luminosity 3$[ab^{-1} ]$ at $\sqrt{s}$ =
		14 TeV.}
	\label{fig:lhctt}
\end{figure}
The figure \ref{fig:lhctt} shows that at 14 TeV, the background is more dominant than the signal. As a result, the signal is not detectable because the cross-section of the triple top is very low even at 14 TeV.
\begin{figure}[!ht]
	\centering
	\includegraphics[width=10cm, height=7cm]{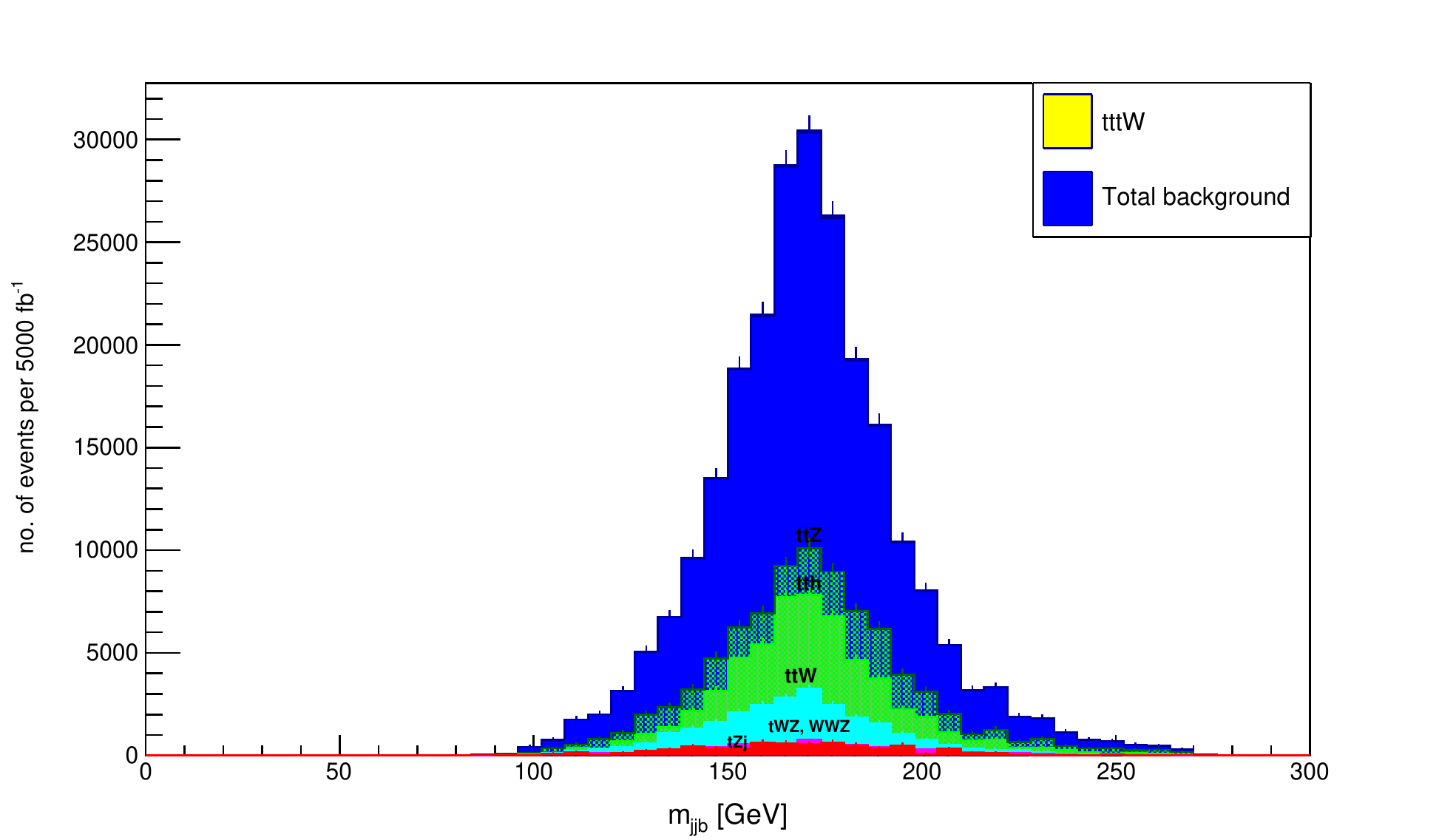}
	\caption{The signal and background samples are normalized to the real number of events obtained at 5000 $fb^{-1}$}
\end{figure}

\begin{table}[ht]
\begin{tabular}{|c|c|c|c|c|c|c|}
\hline
Signal process &\multicolumn{2}{c}{Mass window} & Total & no. of & S/B & Optimized  \\
& Lower limit & Upper limit & Efficiency & Events & &S/$\sqrt{B}$ \\
\hline
$t\bar{t}tW^{+}$ Signal & 186 &294  &0.11  &282  &0.003  & 1.045 \\
 Total Background & 186 & 294 &  & 73143 & & \\
\hline
$t\bar{t}tb$ Signal & 280 & 287 & 0.0002 & 0.03 & 0.0001 & 0.002\\
Total Background & 280 & 287 &  & 325 & & \\
\hline
$t\bar{t}td$ Signal  & 280 & 287 & 0.0003 & 0.12 & 0.0003 & 0.006  \\
 Total Background & 280 & 287 &  & 403 &  & \\
\hline
\end{tabular}
\caption{Signal to background ratio and signal significance values obtained for three different triple top production processes with maximum Integrated luminosity 5000 $fb^{-1}$ at HL-LHC ($\sqrt{s}$=14 TeV) .}
\label{significance1}
\end{table}

\begin{table}[ht]
\begin{tabular}{|c|c|c|c|c|c|c|}
\hline
Signal process &\multicolumn{2}{c}{Mass window} & Total & no. of & S/B & Optimized  \\
& Lower limit & Upper limit & Efficiency & Events & &S/$\sqrt{B}$ \\
\hline
$t\bar{t}tW^{+}$ Signal & 280 &287  &0.0001  &2.27  &0.003  & 0.1 \\
 Total Background & 280 & 287 &  & 628 & & \\
\hline
$t\bar{t}tb$ Signal & 280 & 287 & 0.0003 & 0.22 & 0.0002 & 0.006\\
Total Background & 280 & 287 &  & 1441 & & \\
\hline
$t\bar{t}td$ Signal  & 248 & 296 & 0.0002 & 0.03 & 0.001 & 0.007  \\
 Total Background & 248 & 296 &  & 24 &  & \\
\hline
\end{tabular}
\caption{Signal to background ratio and signal significance values obtained for three different triple top production processes with maximum Integrated luminosity 5000 $fb^{-1}$ at HE-LHC ($\sqrt{s}$=27 TeV) .}
\label{significance}
\end{table}

Because the production process $tt\bar{t}W$ has the largest cross-section for both $\sqrt{s}$ =14 TeV and $\sqrt{s}$ =27 TeV, its S/B ratio is 1. As a result, triple top has observability chances with the production process $tt\bar{t}W$ for both  LHC and HE-LHC. The other two process have very low cross-section and signal significance. 
\section{Conclusion}
In the scope of SM, various channels of triple top quark generation are considered. This analysis of triple-top production shows that it has a very low cross-section when compared to other top-quark production modes. The mass of triple-top is reconstructed using hadronic decay of the top-quark at $\sqrt{s}$=14 TeV. The calculation of signal to background ratio and signal significance clearly shows that at $\sqrt{s}$=14 TeV and $\sqrt{s}$=27 TeV, not all signal scenarios are observable. Only $tt\bar{t}W$ has a signal to background ratio of one. At LHC background are dominant and signal is
not visible because of very low cross-section. Due to the low SM rate, it is conceivable to explore for contributions that are beyond the scope of SM (BSM), which may improve the cross section. Specific computations, however, are required to study actual BSM aspects.

\end{document}